\documentclass[aps,prd,twocolumn,nofootinbib,showpacs]{revtex4}

\usepackage{booktabs,makecell,multirow}
\usepackage{ulem}

\usepackage{graphicx}

\usepackage[colorlinks=true, linkcolor=blue, citecolor=blue, urlcolor=blue]{hyperref}

\begin{document}

\title{The heavy quarkonium inclusive decays using the principle of maximum conformality}

\author{Qing Yu}
\email{yuq@cqu.edu.cn}
\author{Xing-Gang Wu}
\email{wuxg@cqu.edu.cn}
\author{Jun Zeng}
\email{zengj@cqu.edu.cn}
\author{Xu-Dong Huang}
\email{hxud@cqu.eud.cn}
\author{Huai-Min Yu}
\email{yuhm@cqu.edu.cn}

\affiliation{Department of Physics, Chongqing University, Chongqing 401331, People's Republic of China}

\date{\today}

\begin{abstract}

The next-to-next-to-leading order (NNLO) pQCD correction to the inclusive decays of the heavy quarkonium $\eta_Q$ ($Q$ being $c$ or $b$) has been done in the literature within the framework of nonrelativistic QCD. One may observe that the NNLO decay width still has large conventional renormalization scale dependence due to its weaker pQCD convergence, e.g. about $(^{+4\%}_{-34\%})$ for $\eta_c$ and $(^{+0.0}_{-9\%})$ for $\eta_b$, by varying the scale within the range of $[m_Q, 4m_Q]$. The principle of maximum conformality (PMC) provides a systematic way to fix the $\alpha_s$-running behavior of the process, which satisfies the requirements of renormalization group invariance and eliminates the conventional renormalization scheme and scale ambiguities. Using the PMC single-scale method, we show that the resultant PMC conformal series is renormalization scale independent, and the precision of the $\eta_Q$ inclusive decay width can be greatly improved. Taking the relativistic correction $\mathcal{O}(\alpha_{s}v^2)$ into consideration, the ratios of the $\eta_{Q}$ decays to light hadrons or $\gamma\gamma$ are: $R^{\rm NNLO}_{\eta_c}|_{\rm{PMC}}=(3.93^{+0.26}_{-0.24})\times10^3$ and $R^{\rm NNLO}_{\eta_b}|_{\rm{PMC}}=(22.85^{+0.90}_{-0.87})\times10^3$, respectively. Here the errors are for $\Delta\alpha_s(M_Z) = \pm0.0011$. As a step forward, by applying the Pad$\acute{e}$ approximation approach (PAA) over the PMC conformal series, we obtain approximate NNNLO predictions for those two ratios, e.g. $R^{\rm NNNLO}_{\eta_c}|_{\rm{PAA+PMC}} =(5.66^{+0.65}_{-0.55})\times10^3$ and $R^{\rm NNNLO}_{\eta_b}|_{\rm{PAA+PMC}}=(26.02^{+1.24}_{-1.17})\times10^3$. The $R^{\rm NNNLO}_{\eta_c}|_{\rm{PAA+PMC}}$ ratio agrees with the latest PDG value $R_{\eta_c}^{\rm{exp}}=(5.3_{-1.4}^{+2.4})\times10^3$, indicating the necessity of a strict calculation of NNNLO terms.


\end{abstract}

\maketitle

\section{Introduction}

The heavy quarkonium, being a common bound state of Quantum Chromodynamics (QCD) which consists of a pair of heavy quark and antiquark, has been continuously studied either experimentally or theoretically. For example, the $\eta_c$ decays to light hadrons and $\gamma\gamma$ have been measured by the BES III detector~\cite{Ablikim:2012zwa}, which also gives the evidence for $\eta_c\to\gamma\gamma$. In year 2018, the Partical Data Group (PDG)~\cite{Tanabashi:2018oca} issued the most recent information for heavy quarkonium from various measurements. The heavy quarkonium processes involve both perturbative and non-perturbative effects, and these processes are important tests of QCD factorization theories.

The nonrelativistic QCD (NRQCD) factorization theory provides us an effective framework to deal with heavy quarkonium processes~\cite{Bodwin:1994jh}, which factorizes the pQCD approximant into the non-perturbative but universal long-distance matrix elements (LDMEs) and the perturbatively calculable short-distance coefficients. More explicitly, the short-distance coefficients can be expressed as a perturbative series over the strong coupling constant ($\alpha_s$) and the relative velocity of the heavy quarkonium ($v$); e.g. the decay width of the heavy quarkonium $\eta_Q$ can be factorized as the following form~\cite{Bodwin:2002hg}
\begin{eqnarray}
\Gamma_{\eta_{Q}}&=& \frac{F_{1}(^{1}S_{0})} {m^2} \langle\eta_{Q}|\mathcal{O}_{1}(^1S_0)|\eta_Q\rangle
\nonumber\\
&&+\frac{G_{1}(^{1}S_{0})} {m^4} \langle\eta_{Q}|\mathcal{P}_{1}(^1S_0)| \eta_Q \rangle + ... ,
\label{gammaQ}
\end{eqnarray}
where $Q=c$ or $b$, respectively, $F_{1}(^{1}S_{0})$ and $G_{1}(^{1}S_{0})$ are short-distance coefficients. The perturbative series is arranged by the velocity scaling rule. The NRQCD matrix elements $\langle\eta_{Q}|\mathcal{O}_{1}(^1S_0)|\eta_Q\rangle$ and $\langle\eta_{Q}|\mathcal{P}_{1}(^1S_0)|\eta_Q\rangle$ refer to the possibility of observing the specific color and angular-momentum state of the $Q \overline Q$ pair, where the 4-fermion operators $\mathcal{P}_{1}$ and $\mathcal{O}_{1}$ produce or annihilate a $Q\overline Q$ pair in the Fock state $|\eta_Q\rangle$.

It is convenient to compare the following ratio such that to avoid/suppress the uncertainties from the bound state parameters, which is defined as
\begin{eqnarray}
R=\frac{\Gamma_{\eta_Q\rightarrow LH}}{\Gamma_{\eta_Q\rightarrow \gamma\gamma}},
\label{rall}
\end{eqnarray}
where $LH$ stands for light hadrons. The next-to-leading-order (NLO) QCD corrections to the perturbative part of $\eta_c\to LH$ and $\eta_c\to\gamma\gamma$ have been done in Refs.~\cite{Barbieri:1979be, Hagiwara:1980nv}, and the relativistic $\mathcal{O}(\alpha_s v^2)$- corrections have also been done in Refs.~\cite{Czarnecki:2001zc, Guo:2011tz, Jia:2011ah}. Recently, the next-to-next-to-leading order (NNLO) QCD corrections, including the relativistic corrections, have been finished by Feng et al.~\cite{Feng:2017hlu} and Brambilla et al.~\cite{Brambilla:2018tyu}, which also show that factorization scale dependence of the $R$-ratio cancels at the NNLO level. Thus we are facing the chance of achieving a more accurate pQCD prediction on the $R$-ratio.

There is still large renormalization scale dependence for the NNLO pQCD approximant of the $R$-ratio under conventional scale-setting approach. That is, conventionally, people adopts the ``guessed" typical momentum flow of the process such as $m_Q$ as the renormalization scale with the purpose of eliminating the large logarithmic terms or minimizing the contributions of the higher-order loop diagrams~\cite{Wu:2013ei, Wu:2019mky}. Such a naive treatment, though conventional, directly violates the renormalization group invariance~\cite{Wu:2014iba} and does not satisfy the self-consistency requirements of the renormalization group~\cite{Brodsky:2012ms}, leading to a scheme-dependent and scale-dependent less reliable pQCD prediction in lower orders.

To eliminate the unwanted scale and scheme ambiguities caused by using the ``guessed" scale, the principle of maximum conformality (PMC)  scale-setting approach has been suggested~\cite{Brodsky:2011ta, Brodsky:2011ig, Mojaza:2012mf, Brodsky:2012rj, Brodsky:2013vpa}. The purpose of PMC is not to find an optimal renormalization scale but to find the effective coupling constant (whose argument is called as the PMC scale) with the help of renormalization group equation (RGE). By using the PMC, the effective coupling constant is fixed by using the $\beta$-terms of the pQCD series, which are arranged by the general degeneracy relations in QCD~\cite{Bi:2015wea} among different perturbative orders. Since the effective coupling is independent to the choice of the renormalization scale, thus solving the conventional scale ambiguity. Furthermore, after using the PMC to fix the $\alpha_s$ behavior, the remaining coefficients of the resultant series match the series of the conformal theory, leading to a renormalization scheme independent prediction. At the present, the PMC approach has been successfully applied for various high-energy processes.

For the PMC multi-scale method~\cite{Brodsky:2011ta, Mojaza:2012mf}, we need to absorb different types of $\{\beta_i\}$-terms into the coupling constant via an order-by-order manner. Different types of $\{\beta_i\}$-terms as determined from the RGE lead to different running behaviors of the coupling constant at different orders, and hence, determine the distinct PMC scales at each perturbative order. The precision of the PMC scale for higher-order terms decreases at higher-and-higher orders due to the less known $\{\beta_i\}$-terms in those higher-order terms. The PMC multi-scale method has thus two kinds of residual scale dependence due to the unknown perturbative terms~\cite{Zheng:2013uja}, which generally suffer from both the $\alpha_s$-power suppression and the exponential suppression, but could be large due to possibly poor pQCD convergence for the perturbative series of either the PMC scale or the pQCD approximant~\cite{Wu:2019mky}~\footnote{The so-called PMC ambiguity in the recent paper~\cite{Chawdhry:2019uuv} is in fact the second kind of residual scale dependence.}.

Lately, in year 2017, the PMC single-scale method~\cite{Shen:2017pdu} has been suggested to suppress those residual scale dependence. The PMC single-scale method replaces the individual PMC scales at each order by an overall scale, which effectively replaces those individual PMC scales derived under the PMC multi-scale method in the sense of a mean value theorem. The PMC single scale can be regarded as the overall effective momentum flow of the process; it shows stability and convergence with increasing order in pQCD via the pQCD approximates. It has been demonstrated that the PMC single-scale prediction is scheme independent up to any fixed order~\cite{Wu:2018cmb}, thus its value satisfies the renormalization group invariance. Moreover, it has been found that the first kind of residual scale dependence still suffers $\alpha_s$ and exponential suppression and the second kind of residual scale dependence can be eliminated by applying the PMC single-scale method; thus the residual scale dependence emerged in PMC multi-scale method does greatly suppressed. In the following, we shall adopt the PMC single-scale method to do our discussions.

The remaining parts of the paper are organized as follows. In Sec.II, we give the calculation technology for the $R$-ratio under the PMC single-scale method. In Sec.III, we give the NNLO numerical results for $\eta_c$ and $\eta_b$ $R$-ratios and the approximate NNNLO predictions. Sec.V is reserved for a summary.

\section{Calculation technology}

The pQCD approximant for the $R$-ratio has been calculated up to NNLO level under the $\overline{\rm{MS}}$-scheme
\begin{eqnarray}
R &=&\frac{F_{1}(^{1}S_{0})/m^2+{G_{1}(^{1}S_{0})\langle v^2 \rangle_Q/m^2}} {F_{\gamma\gamma}(^{1}S_{0})/m^2+{G_{\gamma\gamma}(^{1}S_{0})\langle v^2 \rangle_Q/m^2}} ,
\label{rconv}
\end{eqnarray}
where those short-distance coefficient $F_{1}(^{1}S_{0})$, $G_{1}(^{1}S_{0})$, $F_{\gamma\gamma}(^{1}S_{0})$ and $G_{\gamma\gamma}(^{1}S_{0})$ can be expressed as
\begin{eqnarray}
F_{1}(^{1}S_{0})&=& \frac{\pi C_F \alpha^2_s}{N_C}(1+f_1\frac{\alpha_s}{\pi}+f_2\frac{\alpha^2_s}{\pi^2}),\\
G_{1}(^{1}S_{0})&=&-\frac{4\pi C_F\alpha^2_s}{3N_C}(1+g_1\frac{\alpha_s}{\pi}) ,\\
F_{\gamma\gamma}(^{1}S_{0})&=&2\pi \alpha^2 e_Q^4(1+\frac{-20+\pi^2}{3}\frac{\alpha_s}{\pi}+f_\gamma\frac{\alpha^2_s}{\pi^2}) ,\\
G_{\gamma\gamma}(^{1}S_{0})&=&2\pi \alpha^2 e_Q^4(-\frac{4}{3}+g_\gamma\frac{\alpha_s}{\pi}).
\label{shrotcoefficients1}
\end{eqnarray}
All the coefficients $f_{1}$, $f_2$, $g_1$, $f_\gamma$, and $g_\gamma$ can be read from Refs.\cite{Feng:2017hlu, Feng:2015uha, Guo:2011tz},
\begin{eqnarray}
f_{1}&=&\frac{8}{9}-\frac{2}{3}\ln2+(\frac{199}{18}-\frac{13}{24}\pi^2)C_A+(\frac{\pi^2}{4}-5)C_F\nonumber\\
&&+\frac{11}{2}L-(\frac{8}{9}+\frac{L}{3})n_f ,\\
f_{2}&=&\hat{f_{2}}-\frac{\pi^2}{2}C_A C_F\ln{\frac{\mu^2_\Lambda}{m^2}}-\pi^2C^2_F\ln{\frac{\mu^2_\Lambda}{m^2}}+[\frac{241}{12}\nonumber\\
&&-\frac{11}{2}\ln2+(\frac{6567}{72}-\frac{429}{96}\pi^2)C_A+(\frac{33}{16}\pi^2-\frac{165}{4})\nonumber\\
&&C_F]L +\frac{363}{16}L^2+\{[-\frac{337}{36}+\frac{1}{3}\ln2+(\frac{13}{48}\pi^2\nonumber\\
&&-\frac{199}{36})C_A+(\frac{5}{2}-\frac{\pi^2}{8})C_F]L-\frac{11}{4}L^2\}n_f+(\frac{4}{9}L\nonumber\\
&&+\frac{L^2}{12})n^2_f,\\
g_{1}&=& \frac{41}{36}-\frac{2}{3}\ln2+(\frac{479}{36}-\frac{11}{16}\pi^2)C_A+(-\frac{49}{12}+2\ln2\nonumber\\
&&+\frac{5}{16}\pi^2-\ln{\frac{\mu^2_\Lambda}{m^2}})C_F+\frac{11}{2}L-(\frac{41}{36}+\frac{L}{3})n_f,\\
f_{\gamma}&=& -8.2196C_F+0.731285C_F\sum^{n_L}_{i=1}\frac{e^2_i}{e^2_Q} -(9.58596\nonumber\\
&&+9.8696\ln{\frac{\mu_\Lambda}{m}})C_A C_F-(40.6123\nonumber\\
&&+19.7392\ln{\frac{\mu_\Lambda}{m}})C^2_F-6.96465C_F L\nonumber\\
&&+(0.0203395C_F+0.4221C_F L)n_f,\\
g_{\gamma}&=&\frac{196}{27}-\frac{5}{9}\pi^2-\frac{96}{27}\ln2+\frac{16}{9}\ln{\frac{\mu^2_\Lambda}{m^2}},
\label{shrotcoefficients2}
\end{eqnarray}
the logarithm $L=\ln{\frac{\mu^2_r}{4m^2}}$ and the non-logarithmic constant
\begin{eqnarray}
\hat{f_{2}}&=&-0.799(13)N^2_C-7.4412(5)n_L N_C-3.6482(2)N_C\nonumber\\
&&+0.37581(3)n^2_L+0.56165(5)n_L+32.131(5)\nonumber\\
&&-0.8248(3)\frac{n_L}{N_C}-\frac{0.67105(3)}{N_C}-\frac{9.9475(2)}{N^2_C}.
\label{nonlogconstant}
\end{eqnarray}
For the $SU(N_C)$ color group, $C_A = N_C$ and $C_F =({N^2_C-1})/{(2N_c)}$ with $N_C=3$. The fine structure constant $\alpha=1/137$. The average of the squared velocity $\langle v^2 \rangle|_{Q}$ of the $(Q\bar{Q})$-quarkonium is
\begin{equation}
\langle v^2 \rangle_Q=\frac{\langle \mathcal{P}_1(^1S_0)\rangle}{m^2_Q \langle\mathcal{O}_1(^1S_0)\rangle},
\end{equation}
and we adopt $\langle v^2\rangle_{c}\simeq 0.430/m_c^2$ and $\langle v^2\rangle_{b}=-0.009$~\cite{Bodwin:2007fz, Chung:2010vz}. The factorization scale $\mu_\Lambda=m$ and the active quark flavor $n_f=n_L+n_H$ with $n_H=1$. Here $n_L=3$ for $\eta_c$ decay and $n_L=4$ for $\eta_b$ decay, and $\sum^{n_L}_{i=1} e^2_i/e^2_Q$ sums up the fractional charges of the light flavors involved in those two decays. The fractional charge $e_Q$ equals to $2\over3$ for $\eta_c$ and $-$${1}\over{3}$ for $\eta_b$, respectively.

Before applying the PMC to the $R$-ratio, we first transform the $\overline{\rm MS}$-scheme pQCD series~(\ref{rconv}) into the one under the minimal momentum space subtraction scheme (mMOM)~\cite{Celmaster:1979dm, MOM}. This transformation avoids the confusion of distributing the $n_f$-terms involving the three-gluon or four-gluon vertexes into the $\beta$-terms~\cite{Du:2017lmz, Binger:2006sj, Zeng:2015gha, Zeng:2018jzf}. This transformation can be achieved by using the relation of the running coupling under the mMOM-scheme and the $\overline{\rm MS}$-scheme, e.g.
\begin{eqnarray}
a_s^{\overline{\rm{MS}}}(\mu_r) &=& a_s^{\rm{mMOM}}(\mu_r)[1+4D_1 a_s^{\rm{mMOM}}(\mu_r)+ \nonumber\\
&&{4^2}D_2 a_s^{\rm{mMOM}, 2}(\mu_r) +{4^3}D_3 (a_s^{\rm{mMOM}, 3}(\mu_r))], \nonumber
\end{eqnarray}
where $a_s(\mu_r)$=${\alpha_s(\mu_r)}/{(4\pi)}$, and under the Landau gauge ($\xi=0$), the first three coefficients $D_1$, $D_2$ and $D_3$ are~\cite{vonSmekal:2009ae}
\begin{eqnarray}
D_1&=&-\frac{169}{48}+\frac{5}{18}n_f,\\
D_2&=&-\frac{18941}{2304}+\frac{351}{128}\zeta_3+(\frac{223}{432}+\frac{1}{12}\zeta_3)n_f \nonumber\\
&& +\frac{25}{324} n_f^2,\\
D_3&=&-\frac{1935757}{110592}+\frac{22485}{2048}\zeta_3+\frac{70245}{4096}\zeta_5+(\frac{42539}{20736}-\nonumber\\
&&\frac{17263}{6912}\zeta_3-\frac{145}{36}\zeta_5) n_f+(\frac{13697}{62208}+\frac{29}{432}\zeta_3)n_f^2 \nonumber\\
&&+\frac{125}{5832} n_f^3,
\end{eqnarray}
where $\zeta_3$ and $\zeta_5$ are usual Riemann zeta functions.

Then, Eq.~(\ref{rconv}) can be rewritten as
\begin{widetext}
\begin{eqnarray}
R&=&\sum_{i\ge1}^{3}r_{i}a_s^{{\rm mMOM}, p+i-1}(\mu_r)\nonumber\\
&=& \sum_{i\ge1}^{3}r_{i,0}a_s^{{\rm mMOM}, p+i-1}(\mu_r)+ \nonumber\\
&& \sum_{i\ge1,j\ge1}^{i+j\le3}(-1)^j [(p+i-1)\beta^{\rm mMOM}(a^{\rm mMOM}_s(\mu_r)) a_s^{{\rm mMOM}, p+i-2}(\mu_r)]  r_{i+j,j} \bigtriangleup_i^{(j-1)}(a^{\rm mMOM}_s(\mu_r)),
\label{Rconv}
\end{eqnarray}
\end{widetext}
where $p=2$, $\bigtriangleup_i^{(j-1)}(a^{\rm mMOM}_s(\mu_r))$ are short notations whose explicit forms can be found in Ref.~\cite{Wu:2019mky}, and $r_{i,0}$ are conformal coefficients and $r_{i,j\neq 0}$ are non-conformal coefficients which can be related to the known $r_{i}$ coefficients by using the standard PMC formulae and by using the degenerated relations among different orders~\cite{Mojaza:2012mf, Brodsky:2013vpa}, i.e.
\begin{eqnarray}
r_1 &=& r_{1,0},  \nonumber\\
r_2 &=& r_{2,0} + p r_{2,1} \beta_0,  \nonumber\\
r_3 &=& r_{3,0} + p r_{2,1} \beta_1 + (p+1) r_{3,1} \beta_0 + {p(p+1)\over2} r_{3,2} \beta_0^2, \nonumber\\
&& \!\!\!\!\!\!\!\!\!\!\!\! \cdots .   \nonumber
\label{pmc}
\end{eqnarray}
The usual $\beta$-function is defined as
\begin{equation}
\beta(a_s(\mu_r)) =-\sum_{i=0}^{\infty}\beta_{i}a^{i+2}_s(\mu_r),
\end{equation}
where the $\beta_i$-functions under the $\overline{\rm MS}$-scheme up to five loop level can be found in Refs.\cite{Gross:1973id, Politzer:1973fx, Caswell:1974gg, Tarasov:1980au, Larin:1993tp, vanRitbergen:1997va, Chetyrkin:2004mf, Czakon:2004bu, Baikov:2016tgj}. The $\beta_0$- and $\beta_1$- functions are scheme independent, and the $\beta_{i>1}$-functions for the mMOM-scheme can be related to the $\overline{\rm MS}$-scheme ones via the relation,
\begin{displaymath}
\beta^{\rm mMOM}(a^{\rm mMOM}_s) = \left({{\partial a^{\rm mMOM}_s}}/{{\partial a^{\overline{\rm MS}}_s}}\right) {\beta^{\overline{\rm MS}}}({a^{\overline{\rm MS}}_s}).
\end{displaymath}
The $\beta_i$-functions under the mMOM-scheme up to four loop level can be found in Ref.~\cite{vonSmekal:2009ae}. The coefficients $r_{i,j}$ are functions of the logarithm ${\rm ln}(\mu_r^2/ 4m_Q^2)$. By using the RGE, these coefficients can be expressed as
\begin{eqnarray}
r_{i,j} = \sum_{k=0}^{j} C_j^k \ln^k(\mu_r^2/4m_Q^2) \hat{r}_{i-k,j-k},
\label{rij}
\end{eqnarray}
where the coefficients $\hat{r}_{i,j}=r_{i,j}|_{\mu_r=2m_Q}$ and the combination coefficients $C_j^k={j!}/{k!(j-k)!}$.

Substituting Eq.~(\ref{rij}) into Eq.~(\ref{Rconv}) and by requiring all the RGE-involved non-conformal terms to zero, one can determine the effective running coupling of the process and hence the optimal scale $Q^\star$ of the process, e.g.
\begin{widetext}
\begin{displaymath}
\sum_{i\ge1,j\ge1,0\le k\le j}^{i+j\le3}(-1)^j \ln^k{Q^2_\star\over 4m_Q^2} [(p+i-1) \beta^{\rm mMOM}(a^{\rm mMOM}_s(Q_\star))a_s^{{\rm mMOM},p+i-2}(Q_\star)]C^k_j\bigtriangleup_i^{(j-1)}(a^{\rm mMOM}_s(Q_\star))\hat{r}_{i+j,j}=0.
\end{displaymath}
\end{widetext}
Thus we obtain
\begin{equation}
R|_{\rm PMC} = \sum^{3}_{i\ge1} \hat{r}_{i,0} a_s^{\rm{mMOM},p+i-1}(Q_\star).
\label{htorrpmc}
\end{equation}

If we have known the NNNLO pQCD series, the PMC scale $Q_{\star}$ can be fixed up to next-next-to-leading-log ($\rm NNLL$) accuracy, e.g.
\begin{eqnarray}
\ln{\frac{Q_{\star}^2}{Q^2}}&=&  T _{0} +T_{1} {\alpha_{s}^{\rm{mMOM}}(Q)\over 4\pi}+T_{2} {\alpha_{s}^{\rm{mMOM},2}(Q)\over 4\pi},
\label{htorrpmcscale}
\end{eqnarray}
where
\begin{widetext}
\begin{eqnarray}
T_0 &=& -{\hat{r}_{2,1}\over \hat{r}_{1,0}},  \\
T_1 &=& {(p+1)(\hat{r}_{2,0}\hat{r}_{2,1}-\hat{r}_{1,0}\hat{r}_{3,1})\over p \hat{r}_{1,0}^2}  +{(p+1)(\hat{r}_{2,1}^2-\hat{r}_{1,0}\hat{r}_{3,2})\over 2\hat{r}_{1,0}^2}\beta_0  \\
T_{2}&=&\frac{(p+1)^2 (\hat{r}_{1,0}\hat{r}_{2,0}\hat{r}_{3,1}-\hat{r}^2_{2,0}\hat{r}_{2,1})
+p(p+2)(\hat{r}_{1,0}\hat{r}_{2,1}\hat{r}_{3,0}-\hat{r}^2_{1,0}\hat{r}_{4,1})}{p^2 \hat{r}^3_{1,0}}+\frac{(p+2)(\hat{r}^2_{2,1}-\hat{r}_{1,0}\hat{r}_{3,2})}{2\hat{r}^2_{1,0}}
\beta_1\nonumber\\
&&-\frac{p(p+1)\hat{r}_{2,0}\hat{r}^2_{2,1}+(p+1)^2(\hat{r}_{2,0}\hat{r}^2_{2,1}-
2\hat{r}_{1,0}\hat{r}_{2,1}\hat{r}_{3,1}-\hat{r}_{1,0}\hat{r}_{2,0}\hat{r}_{3,2})
+(p+1)(p+2)\hat{r}^2_{1,0}\hat{r}_{4,2}}{2p\hat{r}^3_{1,0}}\beta_0\nonumber\\
&&+\frac{(p+1)(p+2)(\hat{r}_{1,0}\hat{r}_{2,1}\hat{r}_{3,2}-\hat{r}^2_{1,0}\hat{r}_{4,3})
+(p+1)(1+2p)(\hat{r}_{1,0}\hat{r}_{2,1}\hat{r}_{3,2}-\hat{r}^3_{2,1})}{6\hat{r}^3_{1,0}}\beta^2_0 .
\end{eqnarray}
\end{widetext}
Using the present known NNLO pQCD seriers, we can fix the PMC scale $Q_\star$ only up to the NLL accuracy. One may observe that the effective scale $Q_\star$ is explicitly independent of the choice of the renormalization scale $\mu_r$ at any fixed order, thus there is no renormalization scale ambiguity for the PMC prediction $R|_{\rm PMC}$. Therefore, the precision of the pQCD approximant can be greatly improved by using the PMC.

It is helpful to estimate how the uncalculated higher-order terms contribute to the pQCD series. Many attempts have been tried in the literature, all of which are based on the known perturbative terms. The usual error estimate obtained by varying the scale over a certain range is unreliable, since it only partly estimates the non-conformal contribution but not the conformal one. Moreover, we should point out that predictions using the conventional renormalization scale-dependent pQCD series~(\ref{rconv}) cannot be reliably used for the purpose; It could have negligible net renormalization scale dependence for the whole pQCD approximant by including enough high-order terms due to the large cancellation among the scale-dependent terms at various orders, but the large scale dependence for each perturbative term cannot be eliminated. On the contrary, the PMC predictions are renormalization scheme-and-scale independent, highly precise values at each order can thus be achieved. As has been pointed out by Ref.\cite{Du:2018dma}, by using the renormalization scheme-and-scale independent conformal series (\ref{htorrpmc}), one can reliably predict how the uncalculated NNNLO-terms contribute to the $R$ pQCD series by using the Pad$\acute{e}$ approximation approach (PAA)~\cite{Basdevant:1972fe, Samuel:1992qg, Samuel:1995jc}.

The PAA was introduced for estimating the $(n+1)_{\rm th}$-order coefficient in a given $n_{th}$-order perturbative Talyor series and feasible conjectures on the likely high-order behaviour of the series. It has been previously demonstrated their applicability to the QCD problems with the help of some resummation methods~\cite{Ellis:1997sb, Burrows:1996dk, Ellis:1996zn, Boito:2018rwt}. For the Pad$\acute{e}$ approximation of a general pQCD approximant $\rho_n(Q)$, its $[N/M]$-type form can be defined as
\begin{eqnarray}
\rho^{[N/M]}_n(Q)  &=&  a^p \times \frac{b_0+b_1 a + \cdots + b_N a^N}{1 + c_1 a + \cdots + c_M a^M}   \label{PAAseries0} \\
                    &=& \sum_{i=1}^{n} C_{i} a^{p+i-1} + C_{n+1}\; a^{p+n}+\cdots,
 \label{PAAseries}
\end{eqnarray}
where $p$ is the starting $\alpha_s$-order, and the parameter $M\geq 1$, $N+M+1=n$. For the present case of $R$-ratio, we have $n=3$. There is a one-to-one correspondence between coefficients $C_{i}$ in Eq.~({\ref{PAAseries}) and the conformal coefficients $\hat{r}_{i,0}$ in Eq.~({\ref{htorrpmc}}). The one-order higher coefficient $C_{n+1}$ can be expressed by those coefficients $b_{i\in[0,N]}$ and $c_{j\in[1,M]}$ which can also be related to the known coefficients $C_i$. More explicitly, if $[N/M]=[n-2/1]$, we have
\begin{equation}
C_{n+1}=\frac{C_n^2}{C_{n-1}}; \label{n-2/1}
\end{equation}
if $[N/M]=[n-3/2]$, we have
\begin{eqnarray}
C_{n+1}=\frac{-C_{n-1}^3+2C_{n-2}C_{n-1}C_{n}-C_{n-3}C_{n}^2}{C_{n-2}^2-C_{n-3}C_{n-1}}; \label{n-3/2}
\end{eqnarray}
if  $[N/M]=[n-4/3]$, we have
\begin{eqnarray}
&&C_{n+1}=\{C_{n-2}^4-(3 C_{n-3} C_{n-1}+2 C_{n-4} C_{n}) C_{n-2}^2 \nonumber \\
&&\quad\quad +2 [C_{n-4} C_{n-1}^2+(C_{n-3}^2+C_{n-5} C_{n-1}) C_{n}] C_{n-2} \nonumber \\
&&\quad\quad -C_{n-5} C_{n-1}^3+C_{n-3}^2 C_{n-1}^2+C_{n-4}^2 C_{n}^2 \nonumber \\
&&\quad\quad -C_{n-3} C_{n} (2 C_{n-4} C_{n-1}+C_{n-5} C_{n})\} \nonumber \\
&&\quad\quad / \{C_{n-3}^3-\left(2 C_{n-4} C_{n-2}+C_{n-5} C_{n-1}\right) C_{n-3} \nonumber \\
&&\quad\quad +C_{n-5} C_{n-2}^2+C_{n-4}^2 C_{n-1}\}; \label{n-4/3} {\rm etc.}
\end{eqnarray}

In the following, we adopt the PMC series~(\ref{htorrpmc}) together with the PAA to estimate the NNNLO contribution to the $R$-ratio. It has been found that for the divergent pQCD series, such as the conventional pQCD series which has renormalon divergence (the dominant factor for the $n_{\rm th}$-order coefficient is proportional to $n!\beta_0^n$), the diagonal PAA series is preferable~\cite{Gardi:1996iq, Cvetic:1997qm}; And if the pQCD series is much more convergent, such as the PMC conformal series which is free of renormalon divergence, the preferred PAA series should be consistent with the GM-L method~\cite{Du:2018dma}.

More explicitly, for the present considered $R$-ratio, the predicted magnitude of the NNNLO-term, is
\begin{equation}
R|^{\rm NNNLO}_{\rm PMC} = \left| \frac{2r_{2,0} r_{3,0}}{r_{1,0}}-\frac{r^3_{2,0}}{r^2_{1,0}} \right| a^{\rm mMOM,5}_s(Q_\star) \label{PAAf}
\end{equation}
for the GM-L-like [0/2]-type Pad$\acute{e}$ series; and
\begin{equation}
R|^{\rm NNNLO}_{\rm PMC} = \left| \frac{r^2_{3,0}}{r_{2,0}} \right| a^{\rm mMOM,5}_s(Q_\star). \label{PAAf}
\end{equation}
for the [1/1]-type diagonal Pad$\acute{e}$ series. Different from previous PMC examples, even though the PMC series of $R$-ratio is free of renormalon divergence, we have found that the diagonal [1/1]-type works much better because the conformal coefficients of are rather large even up to the known NNLO level. This indicates that the divergence property of the $R$-ratio is its inner property, since the PMC series is already scheme-and-scale independent. Then in our following numerical calculation, we shall adopt [1/1]-type to estimate the magnitude of the uncalculated NNNLO terms.

\section{Numerical results}

To do the numerical calculation, we take the quark pole mass from PDG \cite{Tanabashi:2018oca}: the $c$-quark pole mass $m_c=1.67$ GeV and the $b$-quark pole mass $m_b=4.78$ GeV.

\subsection{Basic properties up to NNLO level}

\begin{table}[htb]
\begin{center}
\begin{tabular}{  c c c  c  c c c c c }
\hline
 & ~ ~         & ~$n_f=3$~        & ~$n_f=4$~    & ~$n_f=5$~ \\
\hline
& $\Lambda^{(n_f)}_{\rm QCD}$ (GeV) & 0.502$^{+0.024}_{-0.023}$   &   0.474$^{+0.026}_{-0.025}$  & 0.365$^{+0.023}_{-0.022}$ \\
\hline
\end{tabular}
\caption{The determined asymptotic scale $\Lambda^{(n_f)}_{\rm QCD}$ (in unit: GeV) for the mMOM scheme.} \label{lamda}
\end{center}
\end{table}

As mentioned above, we shall calculate the $R$-ratio under the mMOM scheme, and the Landau gauge is adopted~\footnote{A detailed discussion on the gauge dependence of the mMOM scheme before and after applying the PMC is in preparation~\cite{ZengMOM}.}. The asymptotic scale $\Lambda^{n_f}_{\rm QCD}$ for the mMOM scheme is fixed by using the $\alpha_s$-value at the reference point, $\alpha_s(M_Z) =0.1181\pm0.0011$~\cite{Tanabashi:2018oca} and by using the three-loop RGE. The results are presented in Table~\ref{lamda}.

\begin{figure}[htb]
\centering
\includegraphics[width=0.45\textwidth]{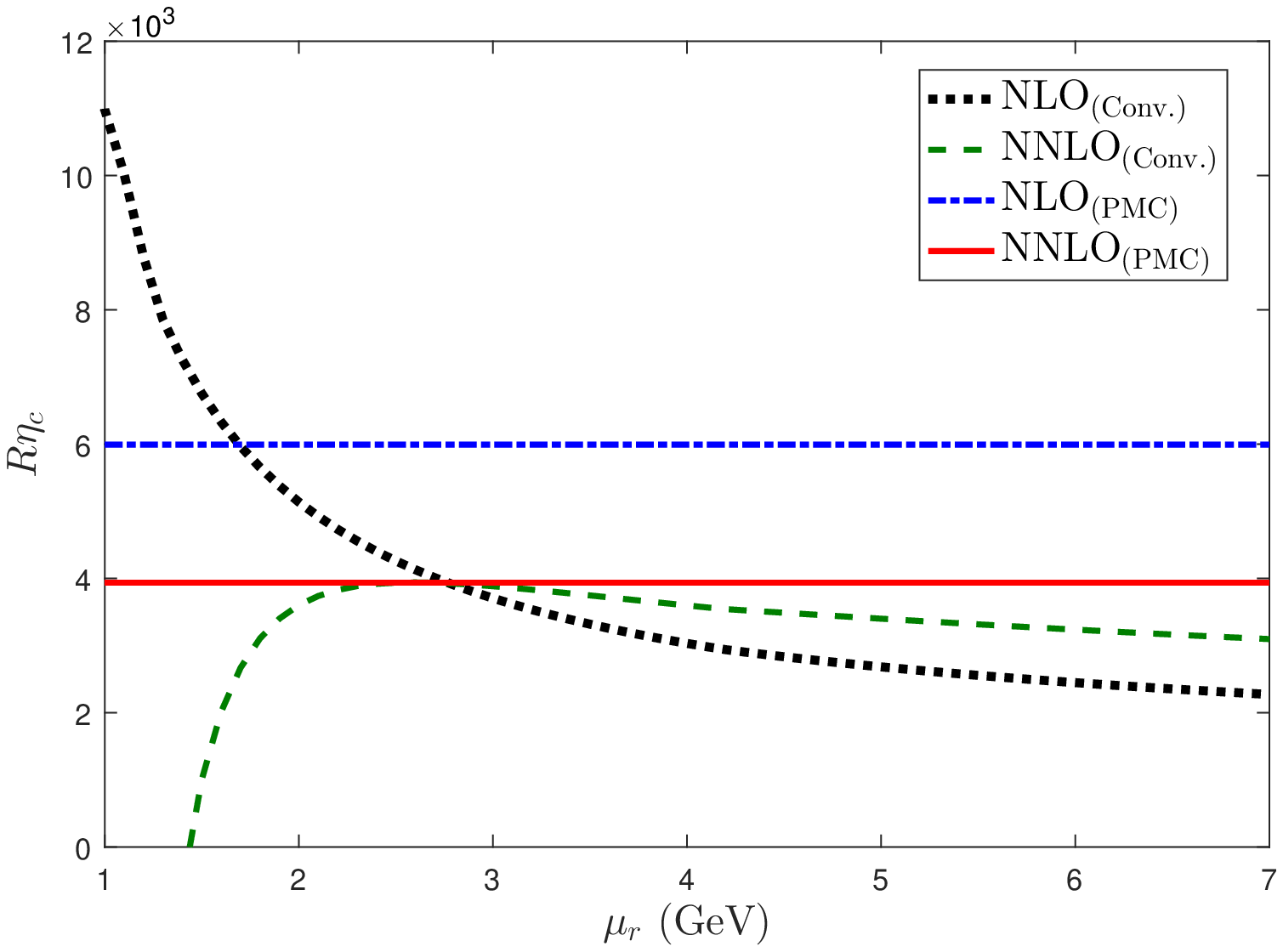}
\caption{The $R_{\eta_c}$-ratio as a function of $\mu_r$ under conventional and PMC scale-setting approaches. The dotted and the dashed lines are for conventional scale-setting up to NLO and NNLO levels, respectively. The dash-dot and the solid lines are for the PMC scale-setting up to NLO and NNLO levels, respectively.}
\label{Plotrc}
\end{figure}

\begin{table}[htb]
\begin{center}
\begin{tabular}{  c c c  c  c c c c c }
\hline
& ~$ R_{\eta_c}(\times 10^3)$~          & ~LO~        & ~NLO~    & ~NNLO~        & ~Total~     \\
\hline
& $\rm{Conv.}$  & 1.87$^{-0.75}_{+2.36}$   &   1.56$^{-0.42}_{+0.28}$  & 0.36$^{+0.33}_{-3.93}$  & $3.79^{+0.15}_{-1.29}$   \\
& $\rm{PMC}$  &   3.54    &   2.45     &     $-$2.06   & 3.93   \\
\hline
\end{tabular}
\caption{Contributions from each loop term for $R_{\eta_c}$-ratios up to NNLO level (in unit: $\times 10^3$) under conventional and PMC scale-setting approaches. The central values are for $\mu_r=2m_c$, and the errors are for $\mu_r\in[m_c , 4 m_c]$. }
\label{etac}
\end{center}
\end{table}

\begin{figure}[htb]
\centering
\includegraphics[width=0.45\textwidth]{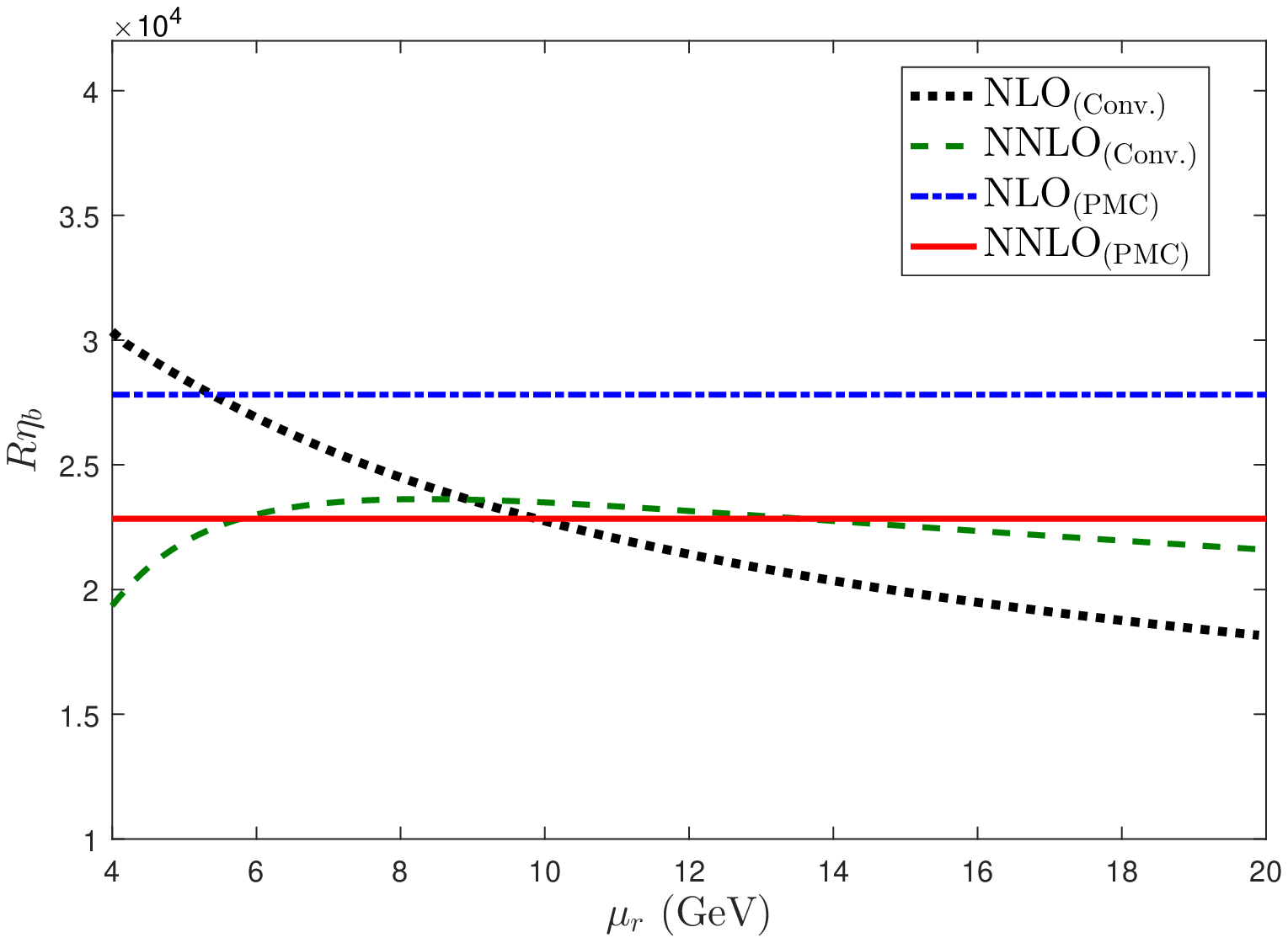}
\caption{The $R_{\eta_b}$-ratio as a function of $\mu_r$ under conventional and PMC scale-setting approaches. The dotted and the dashed lines are for conventional scale-setting up to NLO and NNLO levels, respectively. The dash-dot and the solid lines are for the PMC scale-setting up to NLO and NNLO levels, respectively.}
\label{Plotrb}
\end{figure}

\begin{table}[htb]
\begin{center}
\begin{tabular}{  c c c  c  c c c c c }
\hline
& ~$ R_{\eta_b}(\times 10^3)$~          & ~$\rm{LO}$~        & ~$\rm{NLO}$~    & ~$\rm{NNLO}$~        & ~$\rm{Total}$~     \\
\hline
& $\rm{Conv.}$  & 14.56$^{-4.27}_{+7.76}$   &   8.53$^{-0.42}_{-2.05}$  &  0.46$^{+2.90}_{-7.76}$  & $23.55^{+0.08}_{-2.04}$   \\
& $\rm{PMC}$  &   20.05    &  7.77     &     $-$4.97   & 22.85  \\
\hline
\end{tabular}
\caption{Contributions from each loop term for $R_{\eta_b}$-ratios up to NNLO level (in unit: $\times 10^3$) under conventional and PMC scale-setting approaches. The central values are for $\mu_r=2m_b$, and the errors are for $\mu_r\in[m_b , 4 m_b]$.}
\label{etab}
\end{center}
\end{table}

We present the $R_{\eta_c}$ and $R_{\eta_b}$ ratios up to NNLO level as a function of $\mu_r$ under conventional and PMC scale-setting approaches in Figs.~(\ref{Plotrc}, \ref{Plotrb}). Figs.~(\ref{Plotrc}, \ref{Plotrb}) show that under the conventional scale-setting approach, there are convex behaviors for $R_{\eta_c}$ and $R_{\eta_b}$ ratios within the scale range of $[m_Q, 4m_Q]$, whose peak are $\sim1.6 m_c$ for $R_{\eta_c}$ and $\sim1.7 m_b$ for $R_{\eta_c}$. Contributions from each loop term for $R_{\eta_c}$ and $R_{\eta_b}$ up to NNLO level are presented separately in Tables~\ref{etac} and~\ref{etab}, where the errors are for $\mu_r\in[m_Q, 4m_Q]$. Tables~\ref{etac} and~\ref{etab} show that under conventional scale-setting approach, the renormalization scale dependence for the total NNLO prediction becomes small, whose magnitude is $\sim\left(^{+4\%}_{-34\%}\right)$ for $R_{\eta_c}$, and $\sim\left(^{+0.0}_{-9\%}\right)$ for $R_{\eta_b}$. Such a smaller scale dependence for a NNLO level prediction is caused by the cancellation of scale dependence among different terms, and the scale dependence for each loop term is still very large for conventional series. For example, for the case of $R_{\eta_c}$, the scale errors are $\left(^{-40\%}_{+126\%}\right)$ for the LO-term, $\left(^{-27\%}_{+18\%}\right)$ for the NLO-term, and $\left(^{+92 \%}_{-1092\%}\right)$ for the NNLO-term, respectively. On the other hand, by fixing the effective coupling $\alpha_s(Q_\star)$ of the process with the help of PMC, the renormalization scale dependence for each loop term and hence the total NNLO prediction can be eliminated simultaneously. Thus by applying the PMC, a more accurate pQCD prediction without renormalization scale dependence can be achieved; Basing on the scale-invariant PMC series, it is helpful to predict the unknown NNNLO contribution, as shall be done in next subsection.

Tables~\ref{etac} and~\ref{etab} indicate that under conventional series, even though the NNLO-terms are highly scale dependent, their magnitudes sound more convergent than the PMC series, which are due to accidentally cancellation between the large conformal terms and the divergent non-conformal terms at the NNLO level. This is different from previous PMC examples whose pQCD series converges much more quickly than conventional pQCD series due to the elimination of divergent renormalon terms (and also due to the reason of the magnitude of the conformal terms are usually moderate). By using the PMC, the RGE-involved non-conformal terms have been eliminated and have been adopted to fix the renormalization scale-independent strong coupling constant of the process, $\alpha_s(Q_\star)$. The resultant pQCD series is conformal and scheme independent, one can thus conclude that the PMC series shows the intrinsic property of the pQCD approximant and shows the correct convergent behavior of the pQCD series. At present, the PMC scales for $R_{\eta_{c}}$ and $R_{\eta_{b}}$ can be determined up to NLL-accuracy:
\begin{eqnarray}
\ln{\frac{Q_{\star}^2}{(2 m_c)^2}} &=&  -0.846 -1.005 {\alpha_{s}^{\rm{mMOM}}(2 m_c)} \nonumber\\
                                                        &=& -0.846  -0.299 \nonumber\\
   \Rightarrow \;\;   Q_{\star}   &\simeq& 1.88\;{\rm GeV}
\label{rcpmcscale}
\end{eqnarray}
and
\begin{eqnarray}
\ln{\frac{Q_{\star}^2}{(2 m_b)^2}} &=&  -1.009 -0.280 {\alpha_{s}^{\rm{mMOM}}(2 m_b)} \nonumber\\
                                                        &=&  -1.009 -0.058    \nonumber\\
    \Rightarrow \;\;  Q_{\star}  &\simeq& 5.61\;{\rm GeV},
\label{rbpmcscale}
\end{eqnarray}
which are at the order of ${\cal O}(m_Q)$. The second lines of those two equations show that the second terms are about $35\%$ and $6\%$ of the first terms, indicating the perturbative series of $\ln Q^2_\star/Q^2$ has good convergence at the NLL level for both cases. The effective PMC scale $Q_\star$ is physical, which is renormalization scale independent and determines the correct value of the strong running coupling and hence the correct momentum flow of the process. The heavy quark mass ($m_Q$) provides a natural hard scale for the heavy quarkonium decays into light hadrons or photons, and ${\cal O}(m_Q)$ is usually chosen as the renormalization scale. The PMC scale-setting approach provides a reasonable explanation for this conventional ``guessing" choice.

\subsection{The PAA prediction of the contribution from the uncalculated NNNLO-terms}

Table~\ref{etac} shows that the NNLO PMC prediction of $R_{\eta_c}$ is $3.93\times 10^3$, which is only $\sim65\%$ of the NLO PMC prediction, $R_{\eta_c}=6.09\times 10^3$~\cite{Du:2017lmz}, and is also smaller than the PDG  central value, $R_{\eta_c}^{\rm{exp}}=(5.3_{-1.4}^{+2.4}) \times10^3$~\cite{Tanabashi:2018oca}. This is due to the large negative $\rm{NNLO}$-term~\footnote{A more accurate PMC scale $Q_\star$ at the NLL-accuracy is now achieved by using the NNLO prediction, which reduces the LL-accuracy $Q_\star$ by $\sim14\%$. This leads to an extra difference between the NLO and NNLO predictions.}, as shown by Table~\ref{etac}. Because the pQCD series of $R_{\eta_c}$ shows a slowly convergent behavior, the facts of the previous NLO prediction agrees with the data and the present NNLO prediction does not agree surely do not indicate the failure of the PMC or the pQCD factorization. We should at least know the magnitude of the $\rm{NNNLO}$-term before drawing any definite conclusions.

A strict NNNLO calculation is unavailable in near future due to its complexity. Because the conventional pQCD series cannot be adopted for a reliable prediction, since its known terms are both scale dependent and scheme dependent, and in the following, we shall give an approximation of NNNLO prediction by applying the PAA to the PMC scheme-and-scale independent conformal series.

\begin{table}[htb]
\begin{center}
\begin{tabular}{c c }
\hline
~~~ ~~~ & ~~~NNNLO prediction~~~ \\
\hline
$R_{\eta_{c}}|_{\rm{PMC}} (\times10^3)$    & $5.66^{+0.65}_{-0.55}$  \\
$R_{\eta_{b}}|_{\rm{PMC}} (\times10^3)$   & $26.02^{+1.24}_{-1.17}$  \\
\hline
\end{tabular}
\caption{The predicted NNNLO $R_{\eta_Q}$-ratio by using the [1/1]-type PAA with known NNLO PMC pQCD series, in which the errors are for $\Delta \alpha_s(M_Z^2) = \pm0.0011$. }
\label{pade}
\end{center}
\end{table}

Using the known NNLO PMC conformal series (\ref{htorrpmc}) , the predicted NNNLO $R_{\eta_{c}}$ and $R_{\eta_{b}}$ for [1/1]-type PAA are presented in Table~\ref{pade}. The approximate NNNLO-terms are $1.73\times10^3$ and $3.17\times10^3$ for $R_{\eta_{c}}$ and $R_{\eta_{b}}$, respectively. The absolute values of the LO, NLO, NNLO and the approximate NNNLO terms over the LO-term are $1:0.69:0.58:0.49$ and $1:0.39:0.25:0.16$ for $R_{\eta_{c}}$ and $R_{\eta_{b}}$, respectively. Thus the NNNLO term could still have large contribution and should be taken into consideration for a sound prediction. In fact, Table~\ref{pade} shows that if taking the approximate NNNLO-term into consideration, the $R^{\rm NNNLO}_{\eta_{c}}$-ratio agree well with the PDG value within errors.

\begin{figure}[htb]
\centering
\includegraphics[width=0.48\textwidth]{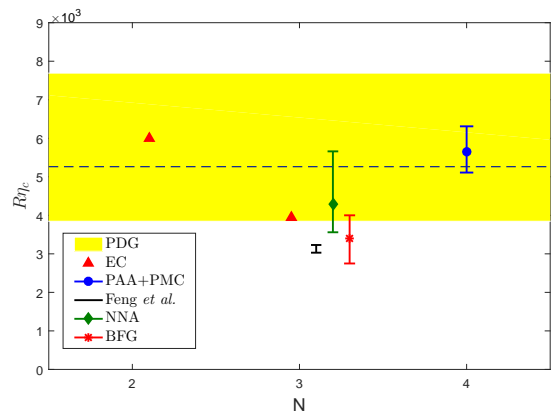}
\caption{The $R_{\eta_c}$-ratio under various approaches. ``EC" is the exact prediction by using the known NLO or NNLO PMC series, and ``PAA+PMC" is the PAA prediction by using the PMC NNLO series. The error of ``PAA+PMC" at the NNNLO level (N=4) is caused by $\Delta \alpha_s(M_Z^2)=\pm0.0011$. As comparisons, the PDG value, the NNLO predictions (N=3) of Feng {\it et al.}~\cite{Feng:2017hlu}, NNA~\cite{Brambilla:2018tyu} and BFG~\cite{Brambilla:2018tyu} are also presented.}
\label{RPAA}
\end{figure}

\begin{figure}[htb]
\centering
\includegraphics[width=0.48\textwidth]{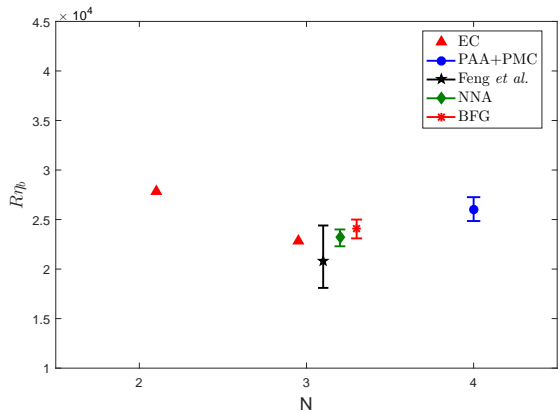}
\caption{The $R_{\eta_b}$-ratio under various approaches. ``EC" is the exact prediction by using the known NLO or NNLO PMC series, and ``PAA+PMC" is the PAA prediction by using the PMC NNLO series. The error of ``PAA+PMC" at the NNNLO level (N=4) is caused by $\Delta \alpha_s(M_Z^2)=\pm0.0011$. As comparisons, the NNLO predictions (N=3) of Feng {\it et al.}~\cite{Feng:2017hlu}, NNA~\cite{Brambilla:2018tyu} and BFG~\cite{Brambilla:2018tyu} are also presented.}
\label{RPAA2}
\end{figure}

In the literature, Ref.\cite{Feng:2017hlu} calculated the $\eta_{c,b}$ decays up to $\rm{NNLO}$ level and gave $R_{\eta_c}=(3.03-3.23)\times10^3$ and $R_{\eta_b}=(20.8^{+3.6}_{-2.7})\times10^3$ by varying $\mu_r$ within the guessed region of 1GeV to $3m_{Q}$. Ref.~\cite{Brambilla:2018tyu} also analyzed the $\eta_{c,b}$ decays up to NNLO level in the large-$n_f$ limit by further using the bubble chain resummation: I) Using the naive non-Abelianization resummation~\cite{Beneke:1994qe}, they obtained $R_{\eta_c}(\rm{NNA})=(4.28^{+1.38}_{-0.72})\times10^3$ and $R_{\eta_b}(\rm{NNA})=(23.2^{+0.8}_{-0.9})\times10^3$; II) Using the background field gauge resummation (BFG)~\cite{DeWitt:1967ub}, they obtained $R_{\eta_c}(\rm{BFG})=(3.39^{+0.61}_{-0.64})\times10^3$ and $R_{\eta_b}(\rm{BFG})=(24.1^{+0.9}_{-1.0})\times10^3$. We should point out that even though those two predictions are consistent with the PDG value, they only partly resum the large renormalon-terms with the purpose of improving the pQCD convergence along~\cite{Brambilla:2018tyu}, which however by using the guessed scale cannot get the correct magnitude of the running coupling and there are still large renormalization scale errors. Thus those predictions cannot be treated as precise pQCD predictions.

Our present PMC predictions are based on the NNLO fixed-order result of Ref.~\cite{Feng:2017hlu}, which includes the important relativistic $\mathcal{O}(\alpha_{s}v^2)$-contribution and has negligible factorization scale dependence. We present the $R_{\eta_c}$-ratio and $R_{\eta_b}$-ratio under various approaches in Figs.~(\ref{RPAA}, \ref{RPAA2}). ``EC" is the exact prediction by using the known NLO or NNLO PMC series, and ``PAA+PMC" is the PAA prediction by using the PMC NNLO series. The PDG value, the NNLO predictions of F.Feng~\cite{Feng:2017hlu}, NNA~\cite{Brambilla:2018tyu} and BFG~\cite{Brambilla:2018tyu} are also presented as comparisons. Different to the previous theoretical predictions whose uncertainties are mainly caused by the renormalization scale dependence, there is no renormalization scale dependence in PMC prediction, and we give a prediction of the error of ``PAA+PMC" approach at the NNNLO level by taking $\Delta \alpha_s(M_Z^2)=\pm0.0011$.

\section{Summary}

In the paper, we have studied the $R_{\eta_Q}$-ratio up to NNLO level. By using the RGE, the scale-independent $\alpha_s$-value can be achieved at any perturbative order, and the correct momentum flow of the process can be determined up to NLL accuracy; and because of the eliminating of the RGE-involved $\{\beta_i\}$-terms, the resultant pQCD series is conformal and scheme independent. Thus we achieve an accurate NNLO fixed-order prediction which is free of conventional renormalization scheme and scale dependence. Fig.\ref{RPAA} shows that the NNLO PMC prediction is somewhat smaller than the PDG value. Table~\ref{etac} shows that the resultant PMC conformal series of $R_{\eta_c}|_{\rm PMC}$ has a slowly convergent behavior, thus before drawing definite conclusions, it is important to make a prediction on the contribution of the uncalculated NNNLO-terms. By using the Pade approximation approach, we give the approximate NNNLO predictions: $R_{\eta_{c}}|_{\rm{PMC}} =5.66^{+0.65}_{-0.55}\times10^3$  and $R_{\eta_{b}}|_{\rm{PMC}} =26.02^{+1.24}_{-1.17}\times10^3$. The approximated NNNLO PMC prediction of $R_{\eta_c}$-ratio agrees with the PDG value within errors, indicating the necessity of finishing a strict NNNLO calculation.

\hspace{1cm}

{\bf\Large Acknowledgments}: This work is supported in part by Natural Science Foundation of China under Grant No. 11625520 and by graduate research and innovation foundation of Chongqing, China under Grant No. CYS19021.

\end{document}